\documentclass[twocolumn,amsmath,amsfonts,amssymb,floats,intlimits]{revtex4}
\usepackage{graphicx}
\def\arcsinh{\mbox{arcsinh}}
\def\div{\mbox{div}}

\begin{document}
\author{D.B. Gutman$^1$,\,  A.D. Mirlin$^{2,3,*}$ \, and \,\,
  Yuval\,\,  Gefen$^4$} 
\affiliation{$^1$The James Franck Institute
5640 S. Ellis Avenue
Chicago, IL 60637 \\
$^2$Institut f\"ur Nanotechnologie, 
Forschungszentrum Karlsruhe,  
76021 Karlsruhe, Germany \\
$^3$Inst. f\"ur Theorie der Kondensierten Materie,
Universit\"at Karlsruhe, 76128 Karlsruhe, Germany\\ 
$^4$Department of Condensed Matter Physics, Weizmann Institute of
Science 76100 Rehovot, Israel}
\title{Kinetic theory of fluctuations in conducting systems}
\date{\today}
\begin{abstract}
We propose  an effective field theory describing  the  
time dependent fluctuations of electrons in conducting systems,
generalizing the well known kinetic theory of fluctuations.
We apply then the theory to analyze
the effects of strong electron-electron and electron-phonon scattering 
on the statistics of current fluctuations. We find 
that if the electron-electron scattering length is much shorter
than the transport mean free path the higher cumulants of current are
parametrically enhanced.

\end{abstract}
\maketitle
Fluctuations of electric current in time
have been the subject of  fundamental and applied research 
since the time of Schottky. They have been a centerpiece in the study 
of   systems ranging  from vacuum tubes to quantum Hall bars 
(for a review see \cite{BB}). More recently noise has been recognized 
as an important diagnostic probe for the nature of 
disorder and interactions.
Traditionally 
the number of electrons taking part  in transport in macroscopic systems
 is large;
for that reason current fluctuations  obey  nearly Gaussian statistics.
Under such circumstances it is the  power  spectrum 
which  is the  object of primary interest  when 
addressing the fluctuations.   
However, as the sample size shrinks  down from a vacuum tube to 
a nanojunction, and the currents  involved become correspondingly smaller,
the central limit theorem does not apply any more,
and information concerning higher current cumulants may be needed.
Various techniques have been applied to address this issue 
\cite{BB,Lesovik,Levitov,TM,Nazarov,GGM}.

There is an  interesting and relevant  class of systems 
out of equilibrium  in which the effect of 
quantum interference on transport quantities is negligible. 
Transport through such  systems
is often  described by classical  kinetic equations  (e.g. the
Boltzmann equation). 
The intensity of the noise  spectrum   can then found employing  
a related approach, known as the  kinetic theory of fluctuations (KTF)
\cite{LL,LP,KS}. 
The appeal of this approach is its   intuitive transparency, yet traditionally
it was   constructed to describe  pair correlation functions, hence 
was limited to finding the  second current cumulant.
In a recent   insightful   work   Nagaev \cite{Nagaev} has proposed
a generalization of   the KTF, expressing  higher order cumulants in
terms of pair correlators. Using this idea (which was termed the
``cascade approach''), Nagaev calculated the third and the fourth
cumulants of the current.  
While this approach was shown to successfully address
a number of problems, its (quantum-mechanical) microscopic
foundations,  
as well as its formulation in terms of an effective field theory
(which would allow for further generalization, such as including
interactions or calculating the full counting statistics (FCS)) 
were not fully understood. An important step in the latter direction  
was an introduction of a stochastic path integral by
Pilgram, Jordan, Sukhorukov, and B\"uttiker \cite{Sukhorukov1}.

In the present  paper we propose  a phenomenological  effective field theory describing  the  
time dependent fluctuations in phase space  of electrons in conducting
systems. 
This is a field-theoretical implementation of the Nagaev's
regression scheme. 
This theory  allows for various generalizations, including  the
effects of electron-phonon and electron-electron interactions.
In a number of instances, this
theory is equivalent (up to small quantum
corrections) \cite{GMG,bagrets} to the non-linear $\sigma$-model
formulation of the problem \cite{GGM}. In particular, for non-interacting
electrons it reproduces the FCS \cite{Lee}
of current fluctuations obtained earlier within quantum approaches.
We apply the theory to systems with a strong electron-electron
scattering (``hot electron regime''). 
We find that (in the absence of umklapp processes), the
current statistics changes dramatically depending on the relation
between the electron-electron scattering length $l_{\rm e-e}$ and the
elastic mean free path. Specifically, in the regime $l_{\rm e-e}\ll
l_{\rm tr}$, when the electrons form a weakly disordered liquid, the
current cumulants are parametrically enhanced due to the
electron-electron scattering. 
We find the scaling law governing the dependence of the FCS 
of hot electrons on the ratio  $l_{\rm e-e}/l_{\rm tr}$.
Finally, we  show that  under strong electron-phonon interaction 
the statistics is Gaussian.

{\it The effective action}. 
The KTF can be cast as a  Boltzmann equation with a 
Langevin-type noise  \cite{KS,Kogan} 
\begin{align}
\label{langevin}
\left({\cal{\hat{L}}}+{\hat{I}}\right)f({\bf p},{\bf x},t)=
\delta J({\bf p},{\bf x},t).
\end{align}
Here we have introduced the  Liouville operator
\begin{align}
{\cal{\hat{L}}}=\frac{\partial}{\partial t}+\frac{{\bf
    p}}{m}\frac{\partial}{\partial {\bf x}}+e{\bf
  E}\frac{\partial}{\partial {\bf p}},   
\end{align}
while $f({\bf p},{\bf x},t) d{\bf x} d{\bf p}$ is the
 number of electrons in a  unit cell centered at 
(${\bf p}, {\bf x}$) in a phase space at time $t$.
With  $\nolinebreak{\rho({\bf x},t)=e\sum_{{\bf p}}f({\bf p},{\bf x},t)}$
  being the local density  of electrons, the self-consistent 
electric field satisfies
$\nolinebreak{\div {\bf E}=4\pi\rho}$.  1 
The operator
$\nolinebreak{\hat{I}}={\hat{I}}_{\rm
 dis}+{\hat{I}}_{\rm e-ph}+\hat{I}_{{\rm e-e}}$ 
is the collision integral \cite{LP} with 
the respective contributions coming from disorder,
electron-phonon and electron-electron scattering. 
The Langevin random source
$\delta J({\bf p},{\bf x},t)$ in (\ref{langevin}) 
represents the  fluctuations of the  net
flux of particle in phase space at  $({\bf p},{\bf x})$. 
These  fluctuations are the consequence of the  scattering processes  
 described by the respective  collision integrals, and should be
 treated consistently 
along with the collision integrals to account for the conserved quantities.
For the physics considered here (and disregarding  quantum
corrections) the interplay among the 
various scattering channels can be  neglected.
Within the same accuracy the  Langevin flux is the sum of 
three uncorrelated random processes  
$\nolinebreak{\delta J=\delta J_{\rm dis}+\delta J_{\rm ee}+\delta
J_{\rm e-ph}}$. 
Its correlation function can be found employing  semiclassical
 considerations \cite{KS}. 
The random flux correlator can be written schematically as 
$\nolinebreak{\langle \delta J({\bf p},{\bf x}, t)\delta J({\bf
 p}',{\bf x}', t)=\delta({\bf x}-{\bf x}')\delta(t-t')
 H[{\bf p},{\bf p}',{\bf x}]}$, where 
$H[{\bf p},{\bf p}',{\bf x}]=H_{\rm dis}[{\bf p},{\bf
 p}',{\bf x}]+H_{\rm ee}[{\bf p},{\bf p}',{\bf x}]+ 
H_{\rm e-ph}[{\bf p},{\bf p}',{\bf x}]$. 
Explicit expressions  for the  collision integrals  and the pair
correlation functions of the  random fluxes are available
\cite{LP,Kogan}.

We consider a
two-terminal system subject to an external voltage bias $V$; the
ambient temperature at the reservoirs is  $T$.
Transport through this system is a stochastic process.
The charge transmitted over a time interval    $\bar{t}$ is
$Q(\bar{t})=\int_0^{\bar{t}} dt' I(t')$  where  $I(t)$
is the electric current. 
The  number of particles $n=Q/e$ is a fluctuating quantity
described by  probability distribution function 
 $P(n)$.
The Fourier transform of the latter 
$\kappa(\lambda)\equiv\sum_ne^{i\lambda n}P(n)$
is  the characteristic function which provides a complete
statistical description of the problem.
Constructing a field theory 
it is natural to look  for a characteristic function  
in a  functional integral representation \cite{GGM}
\begin{align}
\kappa[\tilde{\lambda}]=\int {\cal{D}}f({\bf p},{\bf
  x},t){\cal{D}}\delta \bar{f}({\bf p},{\bf x},t)\exp(iS[f,\delta
\bar{f};\tilde{\lambda}]) . 
\end{align}
Here $f, \delta \bar{f}$ are independent
bosonic fields and the measure of integration is flat.
The field $f({\bf p},{\bf x},t)$ is 
the phase space density  operator within the   KTF;  
 $\tilde{\lambda}({\bf p},{\bf x},t)$ is an auxiliary  'counting field'.
The field $\delta \bar{f}({\bf p},{\bf x},t)$ accounts for fluctuations of this
 'counting field'.
The effective action consists of two  parts    
\begin{align}
\label{total_action}
S[\tilde{\lambda}]=S_{\rm Dyn}+S_{\rm Noise}[\tilde{\lambda}].
\end{align}
The  propagation of fluctuations in  phase space is described by the dynamical 
part of the  effective action:
\begin{align}
\label{effective_action}
iS_{\rm Dyn}=i\int_{\Omega} d{\bf x}\int_{-\infty}^{\infty} dt
{1\over\Omega} \sum_{\bf p} 
\delta\bar{f}({\bf p},{\bf x},t)\left({\cal{\hat{L}}}+\hat{I}\right)
f({\bf p},{\bf x},t).
\end{align}
The second part of the action  accounts for the generation of fluctuations:
\begin{widetext}
\begin{equation}
\label{action_noise}
iS_{\rm Noise}[\tilde{\lambda}]=- \int_{\Omega}d{\bf x}\int_{-\infty}^{\infty}
dt {1\over {\Omega}^2} \sum_{{\bf p},{\bf p}'}  
\left(\tilde{\lambda}({\bf p},{\bf x},t)+\delta \bar{f}({\bf p},{\bf
    x},t)\right) 
H[{\bf p},{\bf p}',{\bf x}]
\left(\tilde{\lambda}({\bf p}',{\bf x},t)+\delta \bar{f}({\bf
    p}',{\bf x},t)\right). 
\end{equation}
The effective action, Eqs.~(\ref{effective_action}) and
(\ref{action_noise}), can be obtained
by exponentiating the equation of motion (\ref{langevin}) with a help of the
Lagrange multiplier $\delta \bar{f}({\bf p},{\bf  x},t)$ and
averaging the resulting $e^{iS}$ over the Langevin fluxes. 
The counting field, as a function of ${\bf p}$, ${\bf x}$, and $t$
should be chosen accordingly  to the problem one is interested in. 
It is illuminating to note that once the  fluctuations  of counting 
field  are neglected in $S_{\rm Noise}$, the effective action, 
(\ref{total_action}) is equivalent to the standard KTF (i.e., it yields
correctly the current-current fluctuations but not  higher cumulants).

So far our treatment has been  general. 
We now assume that the electron gas is degenerate and
consider  a quasi one-dimensional   
diffusive conductor, where  the momentum relaxation length is much
shorter than the system length ($l_{\rm tr}\ll L$).  
The problem then simplifies  considerably.
Additional important parameters are 
the  diffusion coefficient, $D$, the   
single particle density of states $\nu$ at the Fermi level
and the  cross section $S$.
Together they define the Ohmic conductance $G=e^2\nu D S/L$.
Within the diffusive approximation, (\ref{action_noise}) takes the form
\begin{equation}
\label{a7a}
iS_{\rm Dyn}=2\pi\nu i\int _0^Ldx\int_{-\infty}^{\infty} 
d\epsilon\int_{-\infty}^\infty dt
\bigg[\delta\bar{f}(\epsilon,x,t)\bigg\{\frac{\partial}{\partial t}-
D\left(\frac{\partial}{\partial x}
+eE\frac{\partial}{\partial\epsilon}\right)^2 
+{\hat{I}}_{\rm e-ph}+
{\hat{I}}_{\rm e-e}\bigg\}f(\epsilon,x,t)\bigg],
\end{equation}
where $\epsilon=p^2/2m$ is the kinetic energy.
Eq.(\ref{a7a}) describes  propagation of  fluctuations
in diffusive system in  the presence of a long range Coulomb interaction.
In the absence of the Coulomb interaction, one should set $E=0$.

The noise-generating part of the action will also have  three contributions,
corresponding to elastic (impurity), electron-phonon and
electron-electron scattering. 
The noise generated  through  elastic scattering is given by 
\begin{equation}
\label{a7b}
iS_{\rm Noise}^{\rm dis}=-\nu \int dxd\epsilon dt 
D\bigg[\left(\lambda(t) +
eE\frac{\partial \bar{f}(\epsilon,x,t)}{\partial \epsilon}+
\frac{\partial \delta \bar{f}(\epsilon,x,t)}{\partial x}\right)^2 
f(\epsilon,x,t)(1-f(\epsilon,x,t))\bigg].
\end{equation}
\end{widetext}

One can check that the self-consistent field $E$   
does not affect the low frequency current statistics. 
Indeed, a static electric field $E=-\partial_x\phi(x)$ 
can be eliminated from Eqs.~(\ref{a7a}), (\ref{a7b}) by
redefining the energy, $\epsilon + e\phi(x) \to \epsilon$. 
This implies that the counting statistics is the same in the interacting
and non-interacting problems for frequencies much smaller than the
inverse Thouless time, $\omega\ll D/L^2$.
We will consider this regime below and thus set $E=0$.

We are interested in the  statistics of the  number of particles
traversing the  system during   a certain time window,
irrespective  of how this flow  is  distributed over that interval.
We thus select  the counting  field  to be
\begin{align}
\label{a5} \lambda(t)=\left\{
\begin{array}{l}
\lambda, \,\,\,\,\,\,  0<t<\bar{t} \,\, ,
\\
0,  \,\,\,\, {\rm otherwise.}
\end{array}
\right.
\end{align}
Here $\lambda$ is the  argument of the  characteristic function 
$\kappa(\lambda)$.

To analyze the counting statistics 
we use a  saddle point approach, which is justified by a large value
of the dimensionless conductance $g\equiv G/(e^2/h)\gg 1$. Note that
this is the same condition which also
allows us to neglect the quantum corrections to the current statistics
due to the weak localization and the interaction (Altshuler-Aronov)
effects.   
The boundary conditions for the distribution function are
$\nolinebreak{f(\epsilon,0)=f^L(\epsilon), \,f(\epsilon,L)=f^R(\epsilon)}$
where $f^{L(R)}$ are the equilibrium distribution functions of the
electrons on the left (right) reservoir. The auxiliary field  
$\delta\bar{f}$ vanishes at the boundaries.

{\it Current statistics and inelastic processes.}
When the inelastic processes can be neglected (which requires that the 
shortest inelastic length is longer than the sample size,
$l_{\rm in} \gg L$), the saddle-point equations following from
(\ref{a7a}), (\ref{a7b}) permit an analytical solution
\cite{jordan04,bodineau04,GMG}. In particular, in the shot-noise limit 
($eV\gg T$) the resulting action reads  (including a source term for the
average current, $iS^1=iGV\lambda\bar{t}$)
\begin{align}&&
\label{action_elastic}
iS^{\rm dis}=\bar{t}GV\arcsinh^2(\sqrt{e^{i\lambda L}-1}).
\end{align}
The found characteristic function,
$\kappa(\lambda)=\exp\{iS^{\rm dis}\}$
eq.(\ref{action_elastic}),  
agrees with the  known result for the counting
statistics obtained earlier \cite{Lee,Nazarov,GGM} within the 
quantum approaches. 
This demonstrates the equivalence
\cite{GMG}, up to $1/g$ corrections, between our effective theory and the
$\sigma$-model formalism \cite{GGM}.
Indeed, in this case the above effective action
can also be obtained from a particular parametrization of 
non-linear $\sigma$-model \cite{bagrets}.

In the opposite limit, $l_{\rm in} \ll L$, 
the collision integral plays a major role.
There is only a small number of hydrodynamic modes
(determined by the conservation laws) that
propagate for a distances longer that $l_{\rm in}$.
For electron-electron collisions these are the  fluctuations of the
electron density and the energy (or, in other words, of the chemical
potential and the temperature). 
By  the same token, one has to keep only
these two modes in the noise part of the action,  $S_{\rm Noise}$,
Eq.~(\ref{a7b}). 
In the presence of a strong electron-phonon
scattering, $l_{\rm e-ph}\ll L$, the collision integral  $\hat{I}_{\rm
e-ph}$ conserves only the number of electrons. The role
of the electron-phonon scattering 
thus amounts to projecting the
action onto the mode of the density (chemical potential) fluctuations.
 
We consider now  the case of a strong electron-electron scattering,
$l_{\rm ee}\ll L$, in detail. After projecting the fluctuating fields 
$f({\bf p},{\bf x},t)$ and $\delta\bar{f}({\bf p},{\bf x},t)$ in
(\ref{effective_action}), (\ref{action_noise}) onto
the relevant subspace, we are left with the fields $\mu$ and $ T$ 
that parametrize the fluctuations of a chemical potential 
and the temperature, respectively, and the conjugate fields 
$\bar{\mu}$ and $\bar{T}$. 
In terms of these fields the action (\ref{effective_action}),
(\ref{action_noise}) takes the form
\begin{eqnarray}
\label{action_T_mu_Dyn}
&& iS_{\rm Dyn}=i\nu\int dx dt\bigg[\frac{\pi^2}{6}
D^*(T)\partial\bar{T}\partial T^2+D\partial \mu(\partial
\bar{\mu}+\mu\partial \bar{T})\bigg], \nonumber\\
&& \\
\label{action_T_mu_Noise}
&& iS_{\rm Noise}=-\nu\int_0^Ldx\!\int_{-\infty}^{\infty}dt 
T\bigg[D(\lambda+\partial\bar{\mu}\! +
\!\mu\partial\bar{T})^2 \nonumber \\
&& \qquad + D^*(T)\frac{\pi^2}{3}T^2(\partial\bar{T})^2\bigg]. 
\end{eqnarray}
Here we introduced a thermal diffusion coefficient
$\nolinebreak{D^*(T)=v_F^2(\tau^{-1}_{\rm tr}+\tau^{*-1}_{\rm ee}(T))^{-1}/3}$, where
\begin{equation}
\label{tau_ee}
{1\over \tau_{\rm ee}^*(T)}= {3\over \pi^2 \nu T} \int d{\bf p}d{\bf p'} 
{\bf n n'} H_{\rm ee}[{\bf p},{\bf p'}] {\epsilon-\mu\over T}
{\epsilon'-\mu\over T},
\end{equation}
with ${\bf n,n'}$ being the direction of the momenta ${\bf p,p'}$ on
the Fermi surface. Physically, $1/\tau_{\rm ee}^*$ is the relaxation
rate of the energy current due to the electron-electron collision
processes. 
As we are going to show, there are two strikingly different cases.
For $l_{\rm tr} \ll l_{\rm ee}$, which is the commonly considered hot-electron
regime, the thermal diffusion and particle diffusion constants coincide,
$D^*\simeq D$. Our action (\ref{action_T_mu_Dyn}), (\ref{action_T_mu_Noise})
then agrees with that obtained in \cite{pilgram04a,pilgram04b} in the
framework of a stochastic path integral formalism. In this regime, the
electron-electron scattering does not affect the FCS qualitatively but
only modifies the numerical coefficients in the expressions for current
cumulants ${\cal S}_n$. For example, for $eV\gg T$ one finds 
${\cal S}_2=(\sqrt{3}/4)e^2GV$, ${\cal S}_3=(8/\pi^2-9/16)e^3GV$, etc.,
instead of ${\cal S}_2={1\over 3}e^2GV$, 
${\cal S}_3={1\over  15}e^3GV,\ldots$
for non-interacting electrons. 

The situation is qualitatively different in the regime of very strong
electron-electron scattering (or, alternatively, of a very clean system),
$l_{\rm ee} \ll l_{\rm tr}$, when the electrons form a liquid 
flowing through a  weak disorder potential. In the absence of umklapp
processes,  the electron-electron scattering conserves the
total momentum, so that the density diffusion is much faster
than the thermal diffusion, $\alpha^2\equiv D/D^*(T)\gg 1$.
The saddle-point equations following from
(\ref{action_T_mu_Dyn}), (\ref{action_T_mu_Noise}) and determining the FCS of
hot electrons do not permit an analytic solution.
To obtain a qualitative understanding of the problem we first consider
a simplified model, approximating  $D^*(T)$ by a constant $D^*$. 
In this case, the action (\ref{action_T_mu_Dyn}), (\ref{action_T_mu_Noise})
allows us to determine the scaling dependence of the FCS on $\alpha$. 
By performing a rescaling of fields $\bar{\mu}\rightarrow \bar{\mu}/\alpha$, 
$\mu\rightarrow \mu/\alpha$ in  (\ref{action_T_mu_Dyn}),
(\ref{action_T_mu_Noise}) 
we find that $\kappa(\lambda)$ has the following scaling form,
\begin{equation}
\label{fcc_scaling}
\kappa(\lambda)=\exp\{-\alpha^{-2}GT\bar{t}\:{\cal F}(\alpha\lambda,\:\alpha
eV/T)\},
\end{equation}
where the function ${\cal F}(x,y)$ governs the FCS at $\alpha=1$, 
$\kappa(\lambda)=\exp\{-GT\bar{t}\:{\cal F}(\lambda,\:eV/T)\}$. 
This implies that in the equilibrium limit, $T\gg \alpha eV$, the cumulants
${\cal S}_n$ with $n\ge 3$ in the electronic liquid 
are strongly enhanced compared to their values in
the conventional hot-electron regime ($\alpha=1$). Specifically, the
enhancement factor is $\alpha^{2m}$ for the cumulants ${\cal S}_{2m+1}$ and 
${\cal S}_{2m+2}$. In the opposite limit ($T\to 0$ at finite $V$), 
the enhancement starts already from  ${\cal S}_{2}$, with 
${\cal S}_{n}$ enhanced by the factor $\alpha^{n-1}$. Physically, the
enhancement of higher moments of current fluctuations is due to the coupling
between the mode of electric conductivity and the slow mode of the 
thermal conductivity. 

So far we have used a model of a temperature independent thermal diffusion 
coefficient. On the other hand, in the Fermi-liquid theory, 
$D^*(T)\sim\epsilon_Fv_F^2/T^2$.
We argue now that in a realistic situation the above
strong enhancement of the cumulants in the $l_{\rm ee} \ll l_{\rm tr}$ regime
remains parametrically 
valid, with $\alpha^2$ replaced by 
$D^*/D(T_{\rm eff})$, where  $T_{\rm eff}$ is the characteristic
temperature in the system. In particular, in the equilibrium situation
($V=0$), $T_{\rm eff}$ is simply given by the temperatute $T$ of the
reservoirs, $\alpha^2\sim \tau_{\rm tr}/\tau_{\rm ee}(T)$. 
In the non-equilibrium limit
($T=0$, $V$ finite), the effective temperature is found to increase
exponentially with $V$ \cite{Mishchenko}, yielding  asymptotically 
$\alpha^2\sim\exp\{(3/4\pi^2) \tau_{\rm tr}/\tau_{\rm ee}(V)\}$ 

Finally, we consider the case of a strong electron-phonon scattering. 
The only soft modes allowed in this case ($l_{\rm e-ph} \ll L$)
are  fluctuations of chemical potential.
The saddle point equations
$\partial^2\bar{\mu} =0\, ,\,\,\, \partial^2\mu=0 $ 
along  with the boundary  conditions lead to the  trivial solution
\begin{equation}
\label{SP_phonon}
\mu(x)=\mu(0) + (x/L) eV, \qquad
\bar{\mu}(x)= 0.
\end{equation}
The resulting action  
\begin{equation}
iS^{\rm e-ph}=-\bar{t}G[T\lambda^2+iV\lambda]
\end{equation}
represents Gaussian fluctuations, so that higher cumulants are suppressed by
the electron-phonon scattering. 
This is consistent with earlier predictions based on qualitative
arguments \cite{BB} 
and calculations of the third  cumulant \cite{GGM}. However,
to the best of our knowledge, it is the first
analytic derivation of this general result.

To summarize,  we have proposed an effective  theory describing the 
creation and evolution of fluctuations  in phase space
for  degenerate electron gases and liquids. 
We have employed this theory to analyze   
statistical  properties  of noise in disordered conductors.
We have reproduced known results for free fermion and 
confirmed the prediction of Gaussian statistics for the case of strong
electron-phonon scattering. We have analyzed the effect of
electron-electron scattering on the  statistics  of current fluctuations
and established a scaling relation governing the dependence of the FCS on the
ratio $l_{\rm tr}/l_{ee}$, 
showing  that the statistics  differes parametrically  between the
$l_{\rm tr} \ll
l_{ee}$ and $l_{\rm tr} \gg l_{ee}$ cases.  Specifically, in the
latter, when the electrons form a liquid flowing through  weak disorder, 
the current cumulants are
parametrically enhanced by electron-electron scattering. This  is
related to the coupling of the density mode to the slow mode of
temperature fluctuations.  
The regime   $l_{\rm tr} \gg l_{ee}$ is thus  most favorable for
experimental investigation of higher cumulants of current. 

We acknowledge useful discussions and correspondence with 
D.~Bagrets, M.~B\"uttiker, I.~Gruzberg,
Y.~Nazarov, S.~Pilgram, and  P. Wiegmann. 
This work was partially supported by the NSF MRSEC Program under grant
DMR-0213745, by the Minerva Foundation,  the US-Israel BSF ,
 the ISF of the Israel Academy of  Science  and the Alexander von Humboldt foundation (through  the Max-Planck Award ), EC RTN on 
Spintronics. DBG and ADM  acknowledge  support by the
Minerva Einstein Center during their visit to the WIS.

{\it Note added}. As this work was nearing completion we have learned
of several related works
\cite{pilgram04a,pilgram04b,jordan04,bodineau04}
that appeared in the cond-mat preprint archive. 


\end{document}